\documentstyle[aps,preprint,epsbox]{revtex}

\begin{document}
\draft
\preprint{}
\title{Quantum fluctuation induced ordered phase in the Blume-Capel model}

\author{
Norikazu Todoroki$^{1)}$ and Seiji Miyashita$^{2)}$
}
\address{{\rm 1)}Advanced Simulation Technology of Mechanics Co.,Ltd.,
RIKEN, 2-1 Hirosawa, Wako-shi, Saitama, Japan, 351-0198,\\
{\rm 2)}Department of Physics, The University of Tokyo, Bunkyo-ku, Hongo 7-3-1, Tokyo, Japan, 113-8656\\
}

\date{\today}
\maketitle
\begin{abstract}
We consider the Blume-Capel model 
with the quantum tunneling between the excited states.
We find a magnetically ordered phase transition induced by
quantum fluctuation in a model. 
The model has no phase transition in the corresponding classical case.
Usually, quantum fluctuation breaks ordered phase
as in the case of the transverse field Ising model. 
However, in present case, 
an ordered phase is induced by quantum fluctuation.
Moreover, we find a phase transition between 
a quantum paramagnetic phase and a classical diamagnetic phase
at zero temperature.
We study the properties of the phase transition 
by using a mean field approximation (MFA),
and then, by a quantum Monte Carlo method
to confirm the result of the MFA.
\end{abstract}
Recently, in the study of functional material, 
various properties originating from the nontrivial 
energy structure of excited states have attracted much interest. 
In particular, when some cooperative interaction exists 
among the excited states, the system exhibits various 
phases which are incompatible to the ground state, 
and successive phase transitions occur among them. 
For example, the effect of interaction on excitation 
has been studied with regard to mixed-valance biferrocenium salt,\cite{6} and also the photo-induced magnetic properties.\cite{5}

The effect of the excited state is expressed 
by the Blume-Capel model with the variable $S_i^z=0, 
\pm 1$ as the simplest model.\cite{1} 
Here the state $S_i^z=0$ denotes the ground state and the states of 
$S_i^z=\pm 1$ denote the excited states with the magnetization 
or the polarization. In the Blume-Capel model, we assume a cooperative 
interaction among the excited states which could induces ordered states
 even in the cases where the state $S_i^z=0$ has a lower local 
energy because of the kind of crystal field, etc. 
Even if the ordered state is not realized, the interaction 
still causes a metastable state of the magnetized phase 
(or the polarized phase), which could cause interesting 
photo-induced switching phenomena.\cite{5}

In the previous work, we introduced quantum tunneling 
between the states of $S_i^z=\pm 1$ on the Blume-Capel model 
in order to explain the phase transition by isotope replacement 
induced in SrTiO$_3$.\cite{2}
Although the TiO$_6$ cluster of SrTiO$_3$ 
has the six excited states, we considered only the case of 
two excited states, that is, the Blume-Capel model, 
because the structure of the phase diagram does not change 
with the number of excited states. We also considered the 
tunneling only between the excited states because we assumed 
that the tunneling rate between the excited states was higher 
than the tunneling rate between ground state and the excited states
 in SrTiO$_3$.

In this paper, we study the two different effects 
of the quantum tunneling on this model. 
The first one is the effect as the quantum fluctuation. 
That is, as we know in the transverse Ising model, 
\cite{0} the quantum fluctuation suppresses 
the ordered phase, and sometimes it destroys the order 
to cause the quantum disordered phase. We see this effect 
in the present model, too. The second one is the effect 
of the energy gain due to the zero temperature fluctuation. 
That is, the mixing of $S_i^z=\pm 1$ states reduces 
the ground state energy although it also reduces the 
total amount of magnetization. In the case of the ground state 
being in the state $S_i^z=0$, this energy gain may cause 
the magnetized state to have a lower energy. In this case, 
an ordered phase appears in the region where an ordered state 
would not exist when the quantum effect did not exist. 

In this paper we emphasize this second case as a new 
type of quantum effect on the magnetic ordered phase and 
investigate the phase transition. First, we consider the 
mean field approximation (MFA) and next, we perform a quantum 
Monte Carlo simulation to confirm the result of the MFA.

The Hamiltonian of this model is given by 
\begin{eqnarray}
\label{1}
{\cal H}=-J\sum_{\langle i,j\rangle}S_i^zS_j^z
+\Delta \sum_i(S_i^z)^2+\Gamma \sum_i\hat{S}_i^x.
\end{eqnarray}
Here, $S_i^z$ is the $z$-component of the $S=1$ spin operator 
and $\hat{S}_i^x$ is the operator 
which denotes 
the quantum tunneling between 
the excited states:
\begin{eqnarray}
S_i^z
=
\left (
\begin{array}{ccc}
 1& 0& 0 \\
 0& 0& 0 \\
 0& 0&-1 \\
\end{array}
\right ),
\hspace{1cm}
\hat{S}_i^x
=
\left (
\begin{array}{ccc}
 0& 0& 1 \\
 0& 0& 0 \\
 1& 0& 0 \\
\end{array}
\right ).
\end{eqnarray}

First, we treat this model by using 
the MFA. 
We replace $S^z_i$ by $\langle S^z\rangle +\delta_i$ and neglect the higher terms of $\delta$.
Then, the Hamiltonian is given by
\begin{eqnarray}
{\cal H}&=&
\sum_i(-JmzS_i^z+\Delta (S_i^z)^2+\Gamma \hat{S}_i^x )-\frac{1}{2}Jzm^2 \nonumber \\
&=&\sum_i {\cal H}_{\rm MFA}-\frac{1}{2}Jzm^2,
\end{eqnarray}
where
\begin{eqnarray}
\label{meanfieldhamiltonian}
{\cal H}_{\rm MFA}
=
\left (
\begin{array}{ccc}
-Jzm +\Delta -H & 0 & \Gamma \\
                0 & 0 & 0      \\
\Gamma            & 0 & Jzm+\Delta +H
\end{array}
\right ),
\end{eqnarray}
and $z$ denotes the numbers of the nearest neighbor sites
and $H$ is the symmetry breaking field. $m=\langle S^z\rangle$ 
stands for the order parameter.
In this approximation scheme, 
the on-site quantum fluctuation is taken into exactly.
Diagonalizing this Hamiltonian (\ref{meanfieldhamiltonian}),
\begin{eqnarray}
U
\left (
\begin{array}{ccc}
-Jzm +\Delta -H & 0 & \Gamma \\
                0 & 0 & 0      \\
\Gamma            & 0 & Jzm+\Delta +H\\
\end{array}
\right)
U^{-1}
=
\left (
\begin{array}{ccc}
0 &0               & 0               \\
0 &\Delta -\epsilon& 0               \\
0 &0               &\Delta +\epsilon \\
\end{array}
\right ),
\end{eqnarray}
the free energy is obtained as
\begin{eqnarray}
F_{\rm MFA}(m)&=&-k_BT\ln{\rm Tr} \exp \left ( -\beta {\cal H}_{\rm MFA}\right )
-\frac{1}{2}Jzm^2 \nonumber \\
&=&-k_BT \ln (1+2\exp (-\beta\Delta)\cosh (\beta\epsilon ))
-\frac{1}{2}Jzm^2,
\end{eqnarray}
where $\epsilon=\sqrt{(Jzm+H)^2+\Gamma^2}$.

\begin{figure}
\epsfile{file=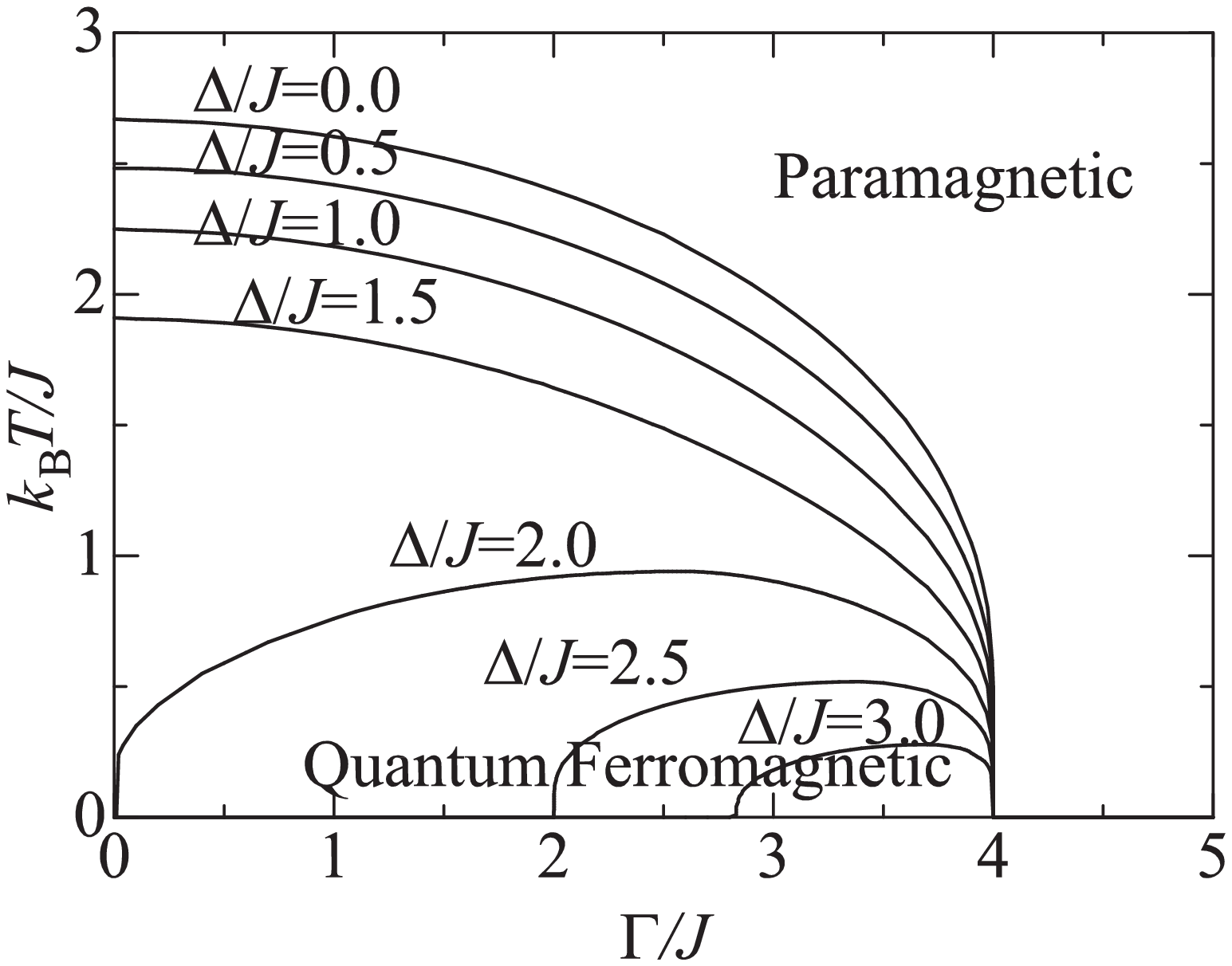,width=10cm}
\caption[]{Phase diagram of the Blume-Capel model with the quantum tunneling between the excited states
in the coordinate ($\Gamma$, $T$). We choose the number of the nearest neighbor sites as $z=4$.\label{fig1}}
\end{figure}

\begin{figure}
\epsfile{file=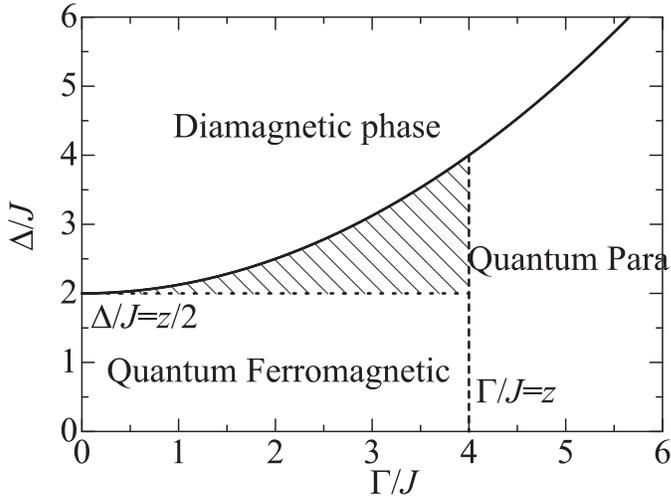,width=10cm}
\caption[]{Ground state phase diagram in the coordinate ($\Gamma$, $\Delta$).
The solid line depict the first order transition 
and the dashed line depict the second order transition. We choose the number of the nearest neighbor sites as $z=4$.\label{fig2}}
\end{figure}

The self-consistent equation for $m$ is 
\begin{eqnarray}
m =\frac{2(Jzm +H)\exp (-\beta\Delta )\sinh (\beta\epsilon )}
{\epsilon (1+2\exp (-\beta\Delta)\cosh (\beta\epsilon ))}.
\end{eqnarray}
We solve this self-consistent equation numerically and we obtain the phase diagram 
as shown in Fig. \ref{fig1}.
In the case of large $\Delta$ ($\Delta>Jz/2$), 
a ferromagnetic phase appears.
This ferromagnetic phase does not exist in the classical model, 
and is induced by quantum fluctuation.
We investigate properties of the ground state in order to clarify the nature of the quantum
fluctuation induced ordered phase.
From the self-consistent equation for $\Gamma < Jz$, 
possible solutions for $m$ are
\begin{eqnarray}
\label{minimaofmag}
m=0, \pm\sqrt{1-\left ( \frac{\Gamma}{Jz}\right ) ^2},
\end{eqnarray}
at $T\rightarrow 0$ and $H\rightarrow 0$. 
For $\Gamma \ge Jz$, the (free) energy has only one minima at $m=0$, 
where the quantum fluctuation destroys the magnetic order. 
For $\Gamma <Jz$, the energies of the system for the solutions (\ref{minimaofmag}) are
\begin{eqnarray}
\begin{array}{lll}
E= 0 && (m=0)\\
E= \Delta-\frac{1}{2}Jz\left (1+\left (\frac{\Gamma}{Jz}\right )^2\right ) &&\left ( m=\pm\sqrt{1-\left ( \frac{\Gamma}{Jz}\right ) ^2}\right )
\end{array}.
\end{eqnarray}
We show the ground state phase diagram in Fig. \ref{fig2}.
When $\Delta$ is small in the region $\Gamma < Jz$, 
the magnetic ordered state of $m\neq 0$ gives the lower energy states than that of $m=0$.
When we increase $\Delta$, the ground state has a quantum phase transition at 
\begin{eqnarray}
\Delta=\frac{1}{2}Jz\left (1+\left (\frac{\Gamma}{Jz}\right )^2\right ),
\end{eqnarray}
which is depicted by a solid curve.
We notice that the critical value of $\Delta$ increases with the
quantum fluctuation. In the classical case,
although the ordered phase does not appear for $\frac{1}{2}Jz <\Delta$.
This phase boundary is indicated by horizontal dotted line.
Therefore, we find that
the ordered phase is induced by the quantum fluctuation 
in the region $\frac{1}{2}Jz <\Delta <Jz$ which is hatched.

The quantum ferromagnetic state $|{\rm Q.F.}\rangle$ is a mixed state of the 
up $|\uparrow\rangle$ and down $|\downarrow\rangle$ states:
\begin{eqnarray}
|{\rm Q.F.}\rangle =p|\uparrow\rangle+q|\downarrow\rangle,
\end{eqnarray}
where $p^2+q^2=1$. For $\Gamma <Jz$,
\begin{eqnarray}
p^2 = \frac{1}{2}\left (1+\sqrt{1-\left (\frac{\Gamma}{Jz}\right )^2}\right ),\hspace{1cm}
q^2 = \frac{1}{2}\left (1-\sqrt{1-\left (\frac{\Gamma}{Jz}\right )^2}\right ),
\end{eqnarray}
and $p^2=q^2=\frac{1}{2}$ for $\Gamma >Jz$.
That is, by the exchange interaction $J$, 
the mixing ratio of the up state and the down state is not the same.
As the result, the quantum ferromagnetic phase appears.
This type of mixed state appears also in the 
transverse field Ising model.
However in the present case, the interesting thing is that 
the quantum fluctuation induces a magnetic order phase 
from the diamagnetic state where $S^z=0$ is dominant.

\begin{figure}
\epsfile{file=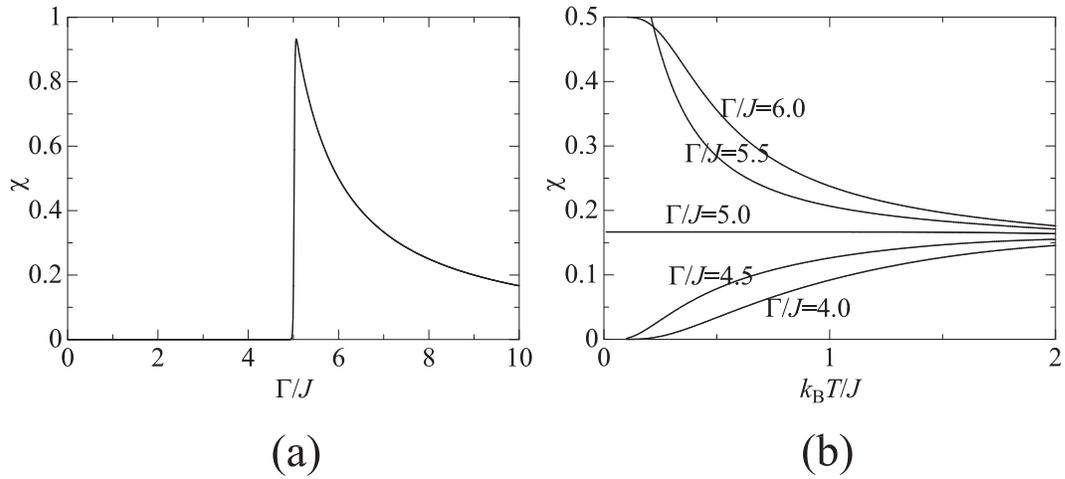,width=15cm}
\caption[]{(a) $\Gamma$ dependence of the susceptibility at $k_BT/J=0.0$ and $\Delta=5.0$. (b) Temperature dependence of the susceptibility at $\Delta =5.0$ and various $\Gamma$.
The transition field is $\Gamma=5.0$.\label{fig3}}
\end{figure}

We also notice that at
the quantum paramagnetic phase is different from the diamagnetic phase,
although the magnetization is zero in both phases.
Therefore, there is a quantum phase transition between these two phases at zero temperature.
Although magnetization is always zero and does not show a jump, 
the susceptibility shows a jump at this quantum phase transition 
as shown in Fig.\ref{fig3}.

\begin{figure}
\epsfile{file=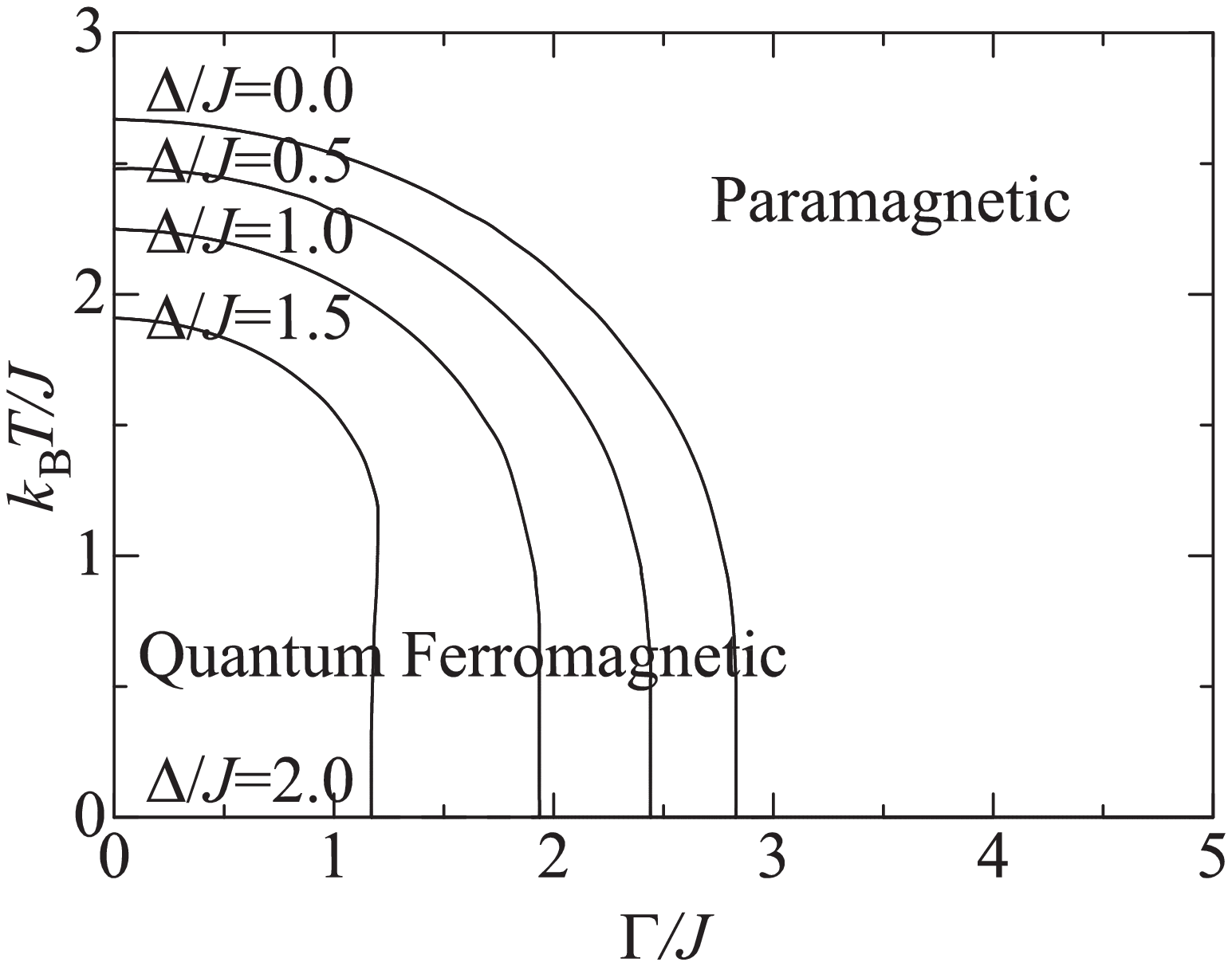,width=10cm}
\caption[]{Phase diagram of the Blume-Capel model with the quantum tunneling between the ground states and the excited states
in the coordinate ($\Gamma$, $T$). We choose the number of the nearest neighbor sites as $z=4$.\label{fig4}}
\end{figure}

In order to realize the present phase diagram,
it is important that 
the tunneling between the excited states is larger than the tunneling 
between the local ground state $S^z=0$ and the local excited states $m=\pm 1$.
For the comparison, we consider a model where the tunneling 
between the ground state and the excited states is given
by the Hamiltonian
\begin{eqnarray}
{\cal H}=-J\sum_{\langle i,j\rangle}S_i^zS_j^z
+\Delta \sum_i(S_i^z)^2+\Gamma \sum_iS_i^x,
\end{eqnarray}
where $S_i^x$ is the $x$ component of $S=1$ spin.
As depicted in Fig. \ref{fig4}, the ordered phase 
is monotonically reduced with $\Gamma$ and ordered phase
induced by the quantum fluctuation
does not appear.

Next, we study properties of the model (\ref{1})
in the square lattice by a quantum Monte Carlo method
with the Suzuki-Trotter decomposition.
In the present model, non-diagonal interaction exists only at each site.
Thus, the model is simply transformed to the classical cubic model.

\begin{figure}
\epsfile{file=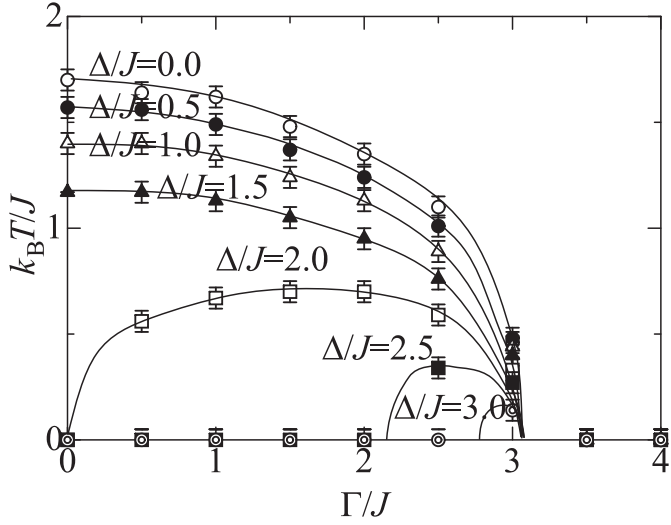,width=10cm}
\caption[]{Phase diagram in the coordinate ($\Gamma$, $T$) by using the quantum Monte Carlo simulation.\label{fig5}}
\end{figure}

\begin{figure}
\epsfile{file=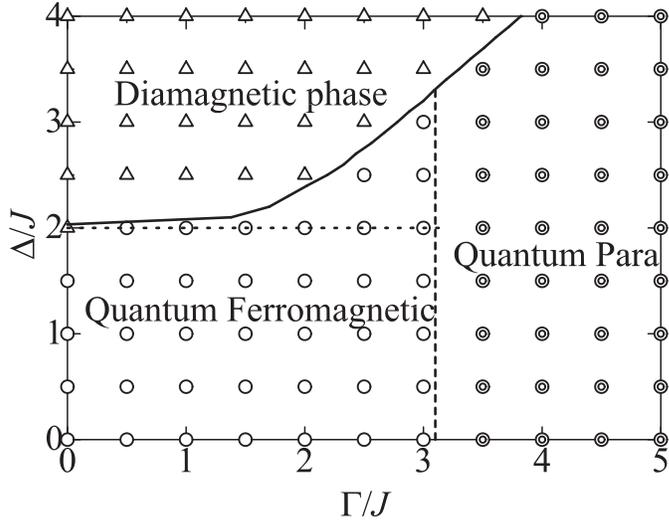,width=10cm}
\caption[]{Ground state phase diagram in the coordinate ($\Gamma$, $\Delta$) by using the quantum Monte Carlo simulation.
The solid line depict the first order transition 
and the dashed line depict the second order transition. \label{fig6}}
\end{figure}

The Suzuki-Trotter decomposition transforms 
the Hamiltonian into a three-dimensional classical model ${\cal H}_{\rm eff}$:
\begin{eqnarray}
{\rm Tr} e^{-\beta{\cal H}}\simeq {\rm Tr}\left ( e^{-\frac{\beta}{N}\left (-J\sum_{\langle i,j\rangle}S_i^zS_j^z+\Delta\sum_i(S_i^z)^2\right )}e^{-\frac{\beta\Gamma}{N}\sum_{i}\hat{S}_i^x}\right )^N \rightarrow {\rm Tr}e^{-\beta{\cal H}_{\rm eff}}.
\end{eqnarray}
${\cal H}_{\rm eff}$ is given with a variable $\sigma =-1,0$, and $1$ as 
\begin{eqnarray}
\beta{\cal H}_{\rm eff}=-K_{N} \sum_{i}\sum_{\mu}\sigma_{i,\mu}\sigma_{i+1,\mu}
         -\tilde{J}\sum_{i}\sum_{\langle \mu,\nu\rangle}\sigma_{i,\mu}\sigma_{i,\nu}
         +\tilde{\Delta}\sum_{i}\sum_{\mu}(\sigma_{i,\mu})^2,
\end{eqnarray}
\begin{eqnarray}
K_N&=&\frac{1}{2}\ln\left (\cot{\frac{\beta \Gamma}{N}}\right ), \\
\tilde{J}&=&\frac{\beta J}{N}, \\
\tilde{\Delta}&=&\frac{\beta \Delta}{N}+\frac{1}{2}\ln \left [\frac{1}{2}\sinh\left (\frac{2\beta \Gamma}{N}\right)\right ].
\end{eqnarray}
Here, $N$ is the Trotter number and
we adopt the periodic boundary condition for the Trotter direction $\sigma_{N+1}=\sigma_1$.
We performed Monte Carlo simulation
by using the cluster heat bath algorithm. \cite{2,4}
In Fig. \ref{fig5}, we present the phase diagram in the coordinate $(\Gamma, T)$.
In Fig. \ref{fig6}, 
we present the ground state phase diagram in the coordinate ($\Gamma$, $\Delta$).
These results are qualitatively same with the results obtained by using the MFA
in the previous section.
Thus, we confirm qualitative features of the result obtained by the MFA.

We found the ordered phase which is induced by the quantum phase transition.
We think that this model is one of the prototype models 
which show the quantum fluctuation induced phase transition.

The numerical calculations were partially performed using the facilities of 
the Supercomputer Center, ISSP, University of Tokyo.
The present work is partially supported by
a Grant-in-Aid from the Ministry of Education, Culture, Sports, Science and Technology.

\end{document}